\documentclass[amssymb,amsmath]{revtex4-1}
\usepackage{graphicx}
\usepackage{amsthm}
\begin{document}

\title{A smoothing monotonic convergent optimal control algorithm for NMR pulse sequence design}

\author{Ivan I. Maximov}
\altaffiliation[Present address: ]{Institute of Neuroscience and Medicine (INM-4), Research Centre J\"{u}lich GmbH, 52425 Germany}
\affiliation{Centre for Insoluble Protein Structures (inSPIN),
Interdisciplinary Nanoscience Centre (iNANO) and
Department of Chemistry, University of Aarhus,
Langelandsgade 140, DK-8000 Aarhus C, Denmark}

\author{Julien Salomon}
\affiliation{CEREMADE, Universite Paris Dauphine, Place de Mareshal de Lattre de Tassigny,
75775 Paris Cedex 16, France}

\author{Gabriel Turinici}
\email{Gabriel.Turinici@dauphine.fr}
\affiliation{CEREMADE, Universite Paris Dauphine, Place de Mareshal de Lattre de Tassigny,
75775 Paris Cedex 16, France}

\author{Niels Chr. Nielsen}
\email{ncn@inano.dk}
\affiliation{Centre for Insoluble Protein Structures (inSPIN),
Interdisciplinary Nanoscience Centre (iNANO) and
Department of Chemistry, University of Aarhus,
Langelandsgade 140, DK-8000 Aarhus C, Denmark}

\date{\today}

\begin{abstract}
The past decade has demonstrated increasing interests in using optimal control based methods within coherent quantum 
controllable systems. The versatility of such methods has been demonstrated with particular elegance within nuclear 
magnetic resonance (NMR) where natural separation between coherent and dissipative spin dynamics processes has 
enabled coherent quantum control over long periods of time to shape the experiment to almost ideal adoption to 
the spin system and external manipulations. This has led to new design principles as well as powerful new experimental 
methods within magnetic resonance imaging,  liquid-state and solid-state NMR spectroscopy. For this development to 
continue and expand, it is crucially important to constantly improve the underlying numerical algorithms to provide 
numerical solutions {which are optimally compatible with implementation on current instrumentation and at same 
time are numerically stable and offer fast monotonic convergence towards the target}. Addressing such aims, we here 
present a smoothing monotonically convergent algorithm for pulse sequence design in magnetic resonance which
{with improved optimization stability} lead to smooth pulse sequence easier to implement experimentally 
and potentially understand within the analytical framework of modern NMR spectroscopy.
\end{abstract}

\pacs{82.56.Jn, 76.60.Fz}
\keywords{Optimal Control Theory, Pulse Sequence Design, NMR, Monotonic Approach}

\maketitle

\section{Introduction}
Optimal control theory (OCT) is a powerful method for control and design of processes within quantum dynamics. 
Originally the method was applied for problems within engineering and economics.\cite{pontryagin,krotov} During
 the past decade or so, optimal-control-based methods have been increasingly used for a development of new 
experiments within optical spectroscopy,\cite{tannor,zhu,rabitz,maday1,ohtsuki,salomon1} quantum information 
processing, \cite{tesch,khaneja1,palao,troppmann,jirari,altafini} liquid- and solid-state nuclear magnetic 
resonance (NMR) spectroscopy, \cite{oc1,gr1,gr2,skin1,oc2,gr3,gr4,gr5,skin2,oc3,skin3,gr6,gr7,oc4,oc5} magnetic 
resonance imaging (MRI),\cite{mri,mri1,mri2,mri3,mri4, mri5,mri6} and dynamic nuclear polarization (DNP) hybrids 
between electron and nuclear magnetic resonance.\cite{dnp1,dnp2,dnp3,dnp4} Such applications have not only been 
useful for the specific disciplines taking advantage of new efficient design procedures and improved experimental 
methods, but it has also stimulated mathematical investigations in quantum optimal control 
theory.\cite{altafini1,altafini2,maday2,maday3,helmke,ito,dalessandro} The latter addresses fundamental questions 
concerned with controllability, convergence, and the establishment of powerful numerical methods for optimal control 
in quantum systems.

So far, the vast majority of optimal control applications within magnetic resonance have taken advantage of gradient-based
 methods, such as the gradient ascent pulse engineering (GRAPE) algorithm introduced for NMR applications by Khaneja and 
coworkers, \cite{oc2} and recently further developed and distributed for general use by Nielsen and coworkers in an 
optimal control version\cite{gr4,oc5} of the open-source NMR simulation software SIMPSON.\cite{simpson1,simpson2} 
In combination with conjugated gradient algorithms, this method \cite{oc2,gr4,oc3,oc5} proves to be  very powerful,
 as demonstrated by numerous applications in which NMR methods with improved  experimental sensitivity, robustness 
towards variations  in instrumental or spin system parameters, reduction of the undesirable effects from dissipative 
processes (i.e., relaxation), and experiments with lower radio-frequency (rf) power requirements and thereby reduced 
risks for sample heating have been developed. More recently, it has been demonstrated \cite{dnp2} that a monotonic 
convergent variant of optimal control method based on the algorithms of Krotov  \cite{krotov,tannor,zhu,maday1} 
represents an interesting alternative to the gradient-based approaches for efficient experiment design. 
Our initial work with this algorithm, in a density operator formulation, exposed important computational 
properties of optimal control methods such as global extremum searching, fast convergence, algorithmic 
simplicity, and independence on time discretization in terms of convergence.

Despite increasing use, it is apparent that current methods face serious challenges in the practical 
realization which needs to be addressed to exploit the full potential of optimal control based methods 
for design of optimal experiments. {This applies not least for the most challenging purposes 
involving large spin systems, optimizations ensuring broadband or band-selective operation with respect to 
certain spin system parameters (e.g., chemical shifts), and powder samples in solid-state NMR spectroscopy. 
For example, looking at the many optimal control pulse sequences proposed so far, it appears that many of these
 display quite wild oscillations in phase and amplitude of the rf control fields  
(see, e.g., Refs.\cite{skin1,skin2,gr3,gr4,gr5,oc3,skin3,gr6,gr7,oc4,oc5})  which may complicate implementation 
on available instrumentation with limitations on the speed and accuracy of phase and amplitude switching.} 
Furthermore, it turns out that GRAPE displays a quite strong dependence on the initial guess of the 
pulse sequence, {dependence on the applied time discretization} (i.e., the number of pulse variables, 
and their duration), as well as unpredictable convergence to local extremum points. Along the same lines, the 
monotonic Krotov-based algorithm faces problems with numerical instability and difficulties in a selection of 
parameters controlling the flow of operations and balance of necessary running costs.\cite{dnp2}  These problems 
have introduced undesired needs for intuition, experience, and repetition of optimizations with a very large set 
of different initial guesses in the usage of these methods as replacement for stronger mathematical recipes.
Although part of these problems have been overpassed in recent adoptions for wave function formalism in optical 
spectroscopy,\cite{maday3} a strong need for solutions to the problems still exists for magnetic resonance 
applications typically performed within a density operator formalism.

In this paper, we present a modified monotonic algorithm which stabilizes convergence and smooths the 
resulting NMR pulses sequences through the use of a frequency truncation technique in course of the optimization. 
The latter  aspect is practically important realizing that most optimal control sequences presented so far 
within NMR spectroscopy display quite significant and fast amplitude and phase modulations, which may cause 
unnecessary problems upon implementation on available spectrometer hardware. This work builds on related 
techniques introduced in the field of the laser control of alignment and rotation~\cite{sugny2} for an unique 
controlling field.

\section{The optimal control problem}

The most typical setup for optimal control pulse sequence design in NMR spectroscopy involves systematic 
generation of optimal radio-frequency (rf) pulse sequences which in a given spin system either (i) accomplish 
the most efficient transfer  of coherence or polarization from a given initial spin state $\rho_0$ to a desired 
target spin state $C$ (often referred to as state-to-state transfer) or (ii) synthesize a specific effective 
(or average) Hamiltonian\cite{haeberlen,hohwy} emphasizing or suppressing certain parts of the internal nuclear 
spin interactions to tailor the Hamiltonian for evolution under desired interactions. These interactions may 
provide a desired state-to-state coherence/polarization transfer, specific information, or provide spectral 
simplification, e.g., in the form of improved spectral resolution.

In absence of dissipative processes, the dynamics of the nuclear spin system may be described by the 
Liouville-von Neumann equation
\begin{eqnarray}
\frac{\mathrm{d}\rho(t)}{\mathrm{d}t} = -i[H(t),\rho(t)] ,
\label{eq1}
\end{eqnarray}
where $\rho(t)$  is the density matrix (initial state: $\rho(0) = \rho_0$) and $H(t)$ is a Hamiltonian of 
the spin system. In the high-field approximation, the latter takes the form
\begin{equation}
H(t)=H_0+\sum_k\omega_k(t)H_k,
\label{eq2}
\end{equation}
with the first term collecting all internal nuclear spin interactions (chemical shifts as well as $J$, 
dipole-dipole, and quadrupole couplings) and the latter describing external rf manipulations with the 
amplitude $\omega_k(t)$ (in angular frequency units) for the spin operator $H_k$ (typically $H_k=I_x,I_y,S_x,S_y$ 
for an I-S two-spin system) being our control fields. The solution to the equation of motion in Eq. (\ref{eq1}) 
is typically expressed as
\begin{equation}
\rho(t)=U(t)\rho_0U^{\dagger}(t),
\label{eq4}
\end{equation}
where the unitary operator (or propagator) $U(t)=\hat D\exp\left( -i\int_0^tH(t^{\prime})\mathrm{d}t^{\prime}\right)$ 
links the unitary evolution with the Hamiltonian in Eq. (\ref{eq2}). $\hat D$ is the so-called Dyson time-ordering operator.

Optimal control relies on optimization of a functional of the type
\begin{equation}
J_i(\omega)=\phi_i-\sum_k\lambda_k\int_0^T\omega^2_k(t)\mathrm{d}t,
\label{eq8}
\end{equation}
with the first term denoting the final cost (or the objective) and the latter term the running cost 
considering the collected energy/power used to reach the optimum. The influence of the running cost is 
scaled by a so-called penalty factor  $\lambda_k$ (which may be constant, as assumed here, or 
time-dependent if, e.g., specific rise and fall time behaviour of the pulses are desired). This convex running 
cost improves the convergence of the optimization methods. Furthermore, it facilitates {development of pulse} 
sequences without too excessive rf power consumption.

For state-to-state transfers between Hermitian operators $C$ and $\rho_0$, the final cost (i.e., the overall 
transfer efficiency) may be expressed as
\begin{equation}
\phi_1=\mathrm{Tr}\left( C\rho(T)\right),
\label{eq5}
\end{equation}
where $T$ is the overall duration of the experiment. For transfers between non-Hermitian operators 
the final cost may be formulated as \cite{oc5,dnp2}
\begin{equation}
\phi_2=|\mathrm{Tr}\left( C^{\dagger}\rho(T)\right)|^2 .
\label{eq6}
\end{equation}
For synthesis of a desired  propagator $U_D$, the final cost is given by
\begin{equation}
\phi_3=\mathrm{Re}\left[\mathrm{Tr}(U(T)U_D^{\dagger})\right].
\label{eq7}
\end{equation}
We note that the objectives $\phi_i$ given above is by no means exclusive, and other variants for the 
target functions may be used. For example, instead of Eq.(\ref{eq7}) one may use the squared expression
 $\phi_3^{\prime}=|\mathrm{Tr}(U_D^{\dagger}U(T)|^2$.
The preferential form of the target function typically depends strongly on the given optimization
 problem. \cite{oc1,oc2,oc3,oc4,oc5}

\section{Monotonic convergence}

The aim of optimal control experiment design is for a given period of time $T$ and a given time discretization
 (i.e., number of pulses, typically taken equidistantly over the time $T$) to find rf control fields 
(i.e., $\omega_k(t)$) which lead to the maximum of the functional in Eq. (\ref{eq8}). The Krotov-based monotonic 
algorithms accomplish this using a Lagrange approach with an adjoint propagator $B(t)$, or a Lagrange multiplier 
for an unconstrained functional as described in Ref. (\cite{dnp2}), using a combination of forward 
propagation with $U(t)$ (Eq.(\ref{eq4}) ) and backward propagation using a conjugated equation for $B(t)$. 
We note that different formulations exist for such calculations, including that of Tannor and 
coworkers \cite{tannor} closely following the original formulations of Krotov, a different variant 
by Zhu et al. \cite{zhu}, and a more general approach (embracing the methods of Tannor and Zhu) as described 
by Maday et al.\cite{maday1}  in context of waveforms and Maximov et al.\cite{dnp2} in context of density operators. 
For simplicity, we here restrict to the original Krotov approach as formulated by Tannor et al.\cite{tannor} 
(corresponding to $\delta=1$ and $\eta=0$ in Refs. (\cite{maday1}) and (\cite{dnp2})), while noting that the more 
general formulation with arbitrary  $\delta$ and $\eta$ parameters may readily be implemented at the expense of
 slightly more complicated formulas.

A prerequisite for monotonic convergence is that the target operator $C$ is positive semi-definite. While this 
condition is obviously fulfilled for optimization of effective Hamiltonians, it will, for example, for optimization 
of transfer between Hermitian operators require the final cost function to be modified to
$\tilde\phi=\phi+\kappa\mathrm{Tr}\left(U(T)U^{\dagger}(T)\right)$ as proposed previously.\cite{dnp2}

The proof for monotonic convergence, in the case of Hermitian operators, may readily be established considering 
the following decomposition of the variation in the cost functional values between arbitrary rf pulses controls 
$\omega$ and
$\omega'$ (in a discrete representation):
\begin{eqnarray}
\nonumber
J(\omega')-J(\omega) &=&
\mathrm{Tr}\left((U'_N-U_N)(U^{'\dagger}_N-U_N^{\dagger})\right)
+\mathrm{Tr}\left(C(U'_N-U_N)\rho_0(U^{'\dagger}_N-U_N)\right)\\
\nonumber
&+&
2\sum_{j=2}^{N+1}\mathrm{Re}\left[\mathrm{Tr}\left\{
\left(e^{i\Delta t\sum_k\omega_{k,j-1}H_k}
e^{-i\Delta t\sum_k\omega'_{k,j-1}H_k}-E\right)AU'_{j-1}B_{j-1}A^{\dagger}\right\}\right]\\
&-&
\Delta t\lambda\sum_k(\omega'_{k,j-1}-\omega_{k,j-1})(\omega'_{k,j-1}+\omega_{k,j-1}),
\label{ap1}
\end{eqnarray}
where we have applied the second-order Strang method to evolve the propagator $U_j$ and the 
Lagrange multiplier $B_j$.\cite{maday3} The time step $\Delta t$ is defined through the number of 
bins $N$ and the overall time  $T=N\Delta t$, while the matrix exponent $A$ is defined as
 $A=e^{-0.5i\Delta tH_0}$. $E$ is the identity operator. We assume all functions to be constant during the 
time step $\Delta t$ rendering computation of the matrix exponents straightforward.

The discrete set of propagators ($U_j$) and Lagrange multipliers ( $B_j$) may be obtained using the following  equations 
\begin{eqnarray}
\nonumber
U_{j+1} &=& Ae^{-i\Delta t \sum_k\omega_{kj}H_k}AU_j,\\
U_0 &=& E, \label{eq9}\\
\nonumber
B_j &=& B_{j+1}Ae^{-i\Delta t \sum_k\omega_{kj}H_k}A,\\
B_N &=& \Theta(C), \label{eq10}
\end{eqnarray}
with the operator $\Theta(C)$ defined by Eqs. (\ref{eq5})--(\ref{eq7}) and the desired efficiency. 
For example, for Hermitian state-to-state transfer 
\begin{equation}
\Theta(C)=\kappa U_N+CU_N\rho_0 ,
\label{ap33}
\end{equation}
where $\kappa$ is an insignificant scaling factor. \cite{dnp2}

To ensure positiveness of the  functional difference $J(\omega')-J(\omega)$ at each time step, we need to 
maximize the last string of Eq.(\ref{ap1}). Following the approach previously described for wave 
functions,\cite{maday3} it appears convenient to re-express the two last terms of Eq.(\ref{ap1}) as a 
function $f_j(\omega'_j)$ depending on a vector variable $\omega'_j$ of dimension $k$ relating to the time step $j$,
\begin{eqnarray}
\nonumber
f_j(\omega'_j)&=&2\mathrm{Re}\left[\mathrm{Tr}\left\{
\left(e^{i\Delta t\sum_k\omega_{k,j-1}H_k}
e^{-i\Delta t\sum_k\omega'_{k,j-1}H_k}-E\right)AU'_{j-1}B_{j-1}A^{\dagger}\right\}\right] \\ 
&-&
\Delta t\lambda\sum_k(\omega'_{k,j-1}-\omega_{k,j-1})(\omega'_{k,j-1}+\omega_{k,j-1}) .
\label{ap2}
\end{eqnarray}

In this formulation, the updated control fields $\omega'_{k,j}$ should be found locally (in time) 
through a minimization problem of dimension $k$ for the vector function $-f_j(\omega^{\prime}_j)$ (corresponding
 to maximization of $f_j(\omega'_j)$ and the corresponding functional in Eq.(\ref{ap1})), which due to the 
non-commutative relationships of the operators $H_k$ in the matrix exponents can not straightforwardly be simplified
 further.\cite{maday1,maday3} The minimization may be conducted starting with an appropriate initial guess, using 
routines such as  conjugated gradients or quasi-Newton.\cite{recipe}  The iteration equation for the rf controls
 $\omega^{\prime}_j$ may be expressed as
\begin{equation}
\omega^{new}_j=\arg\min_{\omega^{\prime}_j}\left\{-f_j(\omega^{\prime}_j)\right\},
\label{ap2_1}
\end{equation}
with the rf fields $\omega_j$ from a previous iteration step used as an initial guess for the minimization problem Eq.(\ref{ap2_1}). 

Equation (\ref{ap2_1}) guarantees monotonic convergence of the functional $J(\omega)$ for all time steps in each 
iteration step through solution of an unconstrained nonlinear optimization problem.  Indeed, $f_j(\omega_j'=\omega_j)=0$, 
so that $-f(\omega_j^{new})\leq 0$. In other words, there exist a pulse sequence 
$\omega_j^{new}$ such that $J(\omega_j^{new})\leq J(\omega_j)$.  Moreover, note that if the algorithm finds 
that  $J(\omega_j^{new}) =J(\omega_j)$ then every local optimization procedure has failed to find a strictly 
better control. In this case, a local maximum has been found.  Note also that the previous approach is independent 
of the optimization method used in the local optimization problems in Eq.(\ref{ap2_1}).

\section{Frequency constraining and smoothing}

Direct application of the monotonic algorithm outlined above, as well as previous gradient-based algorithms, 
often lead to optimal pulse sequences with 
significant oscillations in the rf field amplitudes and phases. {\cite{skin1,skin2,gr3,skin3,oc5,dnp2,dnp4}}
These variations not only hamper analytical understanding of the function of optimal pulse sequences but may 
also complicate practical implementation on available instrumentation. To stimulate generation of smoother 
solutions, we here demonstrate that the optimal control algorithms may be combined with standard frequency 
truncation techniques with a regularization substep that retains monotonic convergence. 

Given $\omega$, an arbitrary control, suppose that a better control $\omega^{new}$ has been obtained, i.e. 
$J(\omega)<J(\omega^{new})$. A way to define a smoother improving control $\omega^{smooth}$ is to consider 
an interpolation between  $\omega^{new}$ (i.e., those typically displaying significant oscillations) and a 
regularized version of it
 (cf. Ref.~\cite{sugny2}).  In this way define for all $k$
\begin{equation}
\omega_k^{\mathrm{smooth}}=(1-\alpha)\omega_k^{\mathrm{new}}+\alpha\mathbb{F}(\omega_k^{\mathrm{new}}),
\label{ap3}
\end{equation}
where $\alpha$ is a regularization parameter and $\mathbb{F}(\omega)$ defines the frequency truncation. 
{The frequency truncation may be accomplished in many ways, e.g., using built-in functions in MATLAB
 \cite{matlab}  or simply by Fourier transforming the pulse sequence, removing high-frequency components, and 
transforming it back to the time domain using an inverse Fourier transformation.}

Following the proofs in the previous section, it is obvious that $\alpha=0$ ensure monotonic behaviour and 
thereby $J(\omega^{\mathrm{smooth}}_{\alpha=0})\geq J(\omega)$. The same is  almost true when $\alpha\rightarrow 0$ 
as well. Accordingly, an iterative sub-algorithm may be established which generates a smooth and monotonic solution:
\begin{enumerate}
\item start from $\alpha=1$;
\item test $J(\omega^{\mathrm{smooth}})\geq J(\omega)$?
\item if not, decrease $\alpha$, for example, halving the value and go to step 2.
\end{enumerate}

With this ingredient and the formula in Eqs. (\ref{eq9}) - (\ref{ap33}), we can formulate the overall algorithm. 
Consider a parameter $Tol>0$.

\begin{enumerate}
\item  Set the initial random guess $\omega^0=(\omega^0_{j,k})$; $j=1,..,N;$ $k>0$.

\item Compute the corresponding state $U_0$ and $B_0$ according to Eqs. (\ref{eq9})-(\ref{eq10}).

\item Set $\epsilon=+\infty$, $\ell=0$.

\item While $\epsilon>Tol$ do

		a. Do forward propagation and search new rf field
$\widetilde{\omega}^{\ell+1}$ according to the procedure in Sec. III

		b. Apply the frequency truncation sub-algorithm with regularization to
find $\alpha^\ell$ that preserves the monotonicity, i.e.:
$J((1-\alpha^\ell)\widetilde{\omega}^{\ell+1}+\alpha^\ell F(\widetilde{\omega}^{\ell+1})) \geq J(\omega^\ell)$

		c. Define
$\omega^{\ell+1}=(1-\alpha^\ell)\widetilde{\omega}^{\ell+1}+\alpha^\ell F(\widetilde{\omega}^{\ell+1})$

		d. Set $\epsilon=J(\omega^{\ell+1})-J(\omega^\ell)$

		e. Do backward propagation to compute $B^{\ell+1}$ the solution of Eq. (\ref{eq10})
with $\omega=\omega^{\ell+1}$

		f. Do $\ell=\ell+1$

  End While.
\end{enumerate}

It is important to note that frequency truncation technique with regularization could equally well be
applied to other known optimal control approaches, such as GRAPE and different variants of Krotov-based optimal control. 

\section{Design of smooth NMR experiments}

While the overall applicability of optimal control for NMR experiment design has been demonstrated and verified 
numerically and experimentally in numerous previous papers,\cite{oc1,gr1,gr2,oc2,gr3,gr4,gr5,oc3,gr6,gr7,oc4,oc5} 
 we will demonstrate numerically in this paper the implication of our smoothing procedure and monotonic convergence 
in optimal control design of a set of simple NMR experiments. We address specifically spin-state selective, 
non-Hermitian coherence transfer through $J$ couplings for liquid-state NMR applications and excitation of the central 
transition of $^{23}$Na for applications in magnetic resonance imaging (MRI). {Lists containing  rf 
amplitudes and phases for the optimal control pulse sequences as well as Matlab scripts used to generate these can be found in Supplementary Material.\cite{epaps}}

\subsection{Coherence-order and spin-state-selective coherence transfer}

For a two-spin, $J$-coupled spin systems in the context of liquid-state NMR, the internal Hamiltonian may in 
the high-field approximation be cast as
\begin{equation}
H_0=\pi J 2 I_zS_z,
\label{eq12}
\end{equation}
where  $I_z$ or $S_z$ represent $z$-components of the spin operators for the $I$ and $S$ spins in the present 
example coupled through a scalar coupling of size $J$ = 140 Hz. To exemplify transfer between non-Hermitian 
operators, we assume the initial state to be represented by +1-quantum coherence on the $S$ spin (i.e.,  $\rho_0=S^+$) 
and to further demonstrate spin-state-selective transfer,\cite{s31,s32,s33,s34} we assume that the destination operator 
is -1-quantum coherence on the $I$ spin with the $S$ spin being in the $\alpha$-state corresponding to only one of the 
lines in the $J$-coupled doublet excite (i.e., $C$=$I^-S^{\alpha}$). 

In this case of transfer between non-Hermitian operators, we use the target function in Eq.(\ref{eq6}) 
modified to ensure {our target} being positive semi-definite. This leads to the modified  objective 
$\tilde\phi$ and the operator $\Theta(C)$: 
\begin{eqnarray}
\tilde\phi &=& \phi_2+\mathrm{Tr}(U(T)U^{\dagger}(T)),\\
\nonumber
\Theta(C) &=& U^{\dagger}(T)+\rho_0U^{\dagger}(T)C^{\dagger}\mathrm{Tr}\left[U^{\dagger}(T)\rho_0^{\dagger}U(T)C\right]\\
&+&
C^{\dagger}U^{\dagger}\rho_0\mathrm{Tr}\left[U(T)\rho_0U^{\dagger}(T)C^{\dagger}\right],
\end{eqnarray}
where we, relative to the expression in Eq. (\ref{ap33}), assumed the scaling factor to be $\kappa\equiv 1$ 
for the sake of simplicity. Our control fields are represented by the amplitudes (angular frequencies)  
$\omega_k(t)$ of $x$-  and $y$- phase rf irradiation on the spins $I$ (operators $I_x$ ($k$=1) and $I_y$ ($k$=2)) 
and $S$ (operators $S_x$ ($k=3$) and $S_y$ ($k=4$)). According to unitary bounds on spin dynamics,\cite{stoustrup,science}
 the maximum achievable transfer efficiency in this case is 1. For the optimizations we set the excitation 
period to $T$ = 7.14 ms  (corresponding to 1/$J$) and the number of pulses to $N$ = 200. 

\begin{figure}
\includegraphics[scale=0.35,angle=-90]{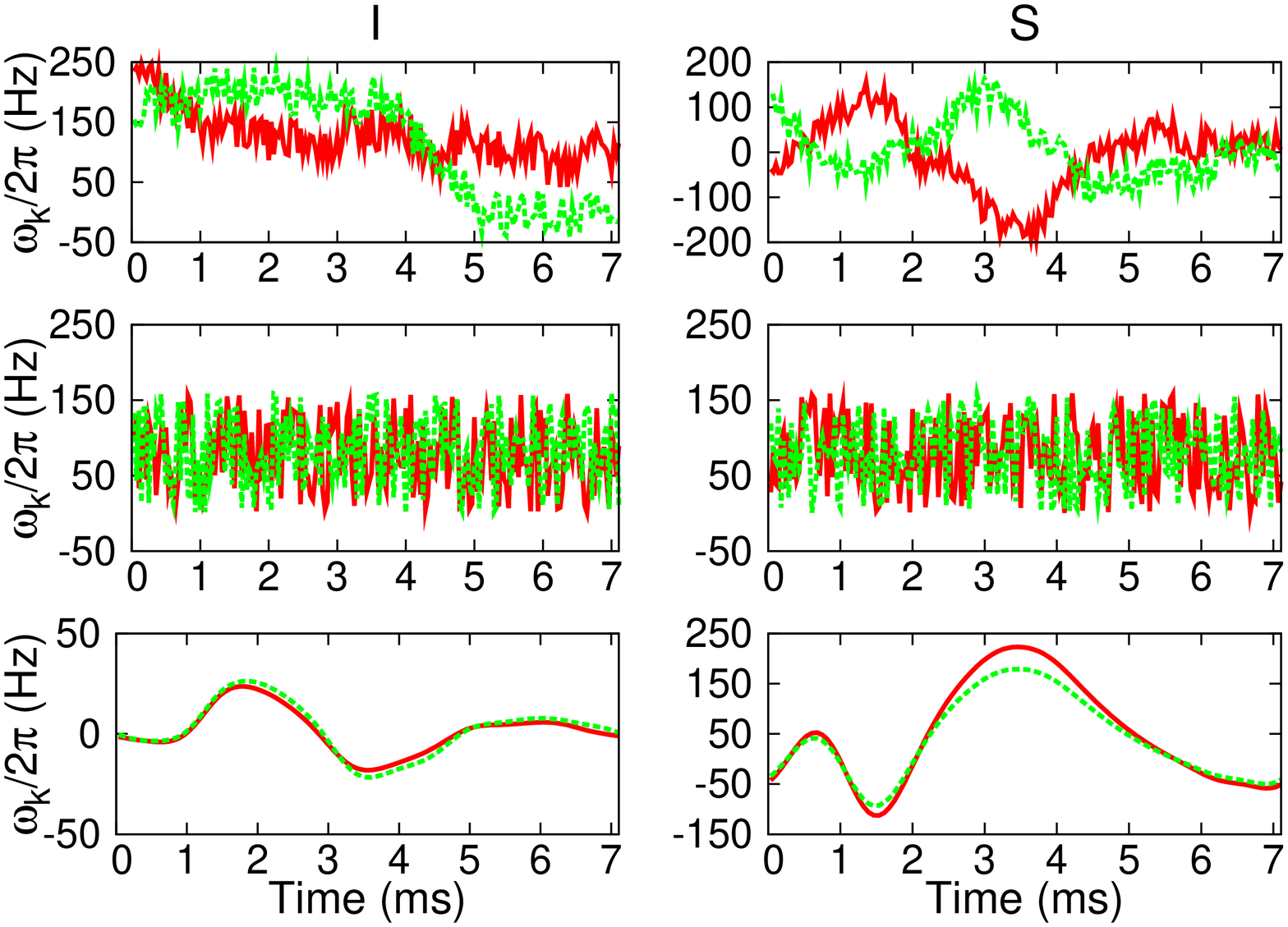}
 \caption{}
\label{fig1}
\end{figure}

Addressing this specific optimization problem, Figure \ref{fig1} compares pulse sequences (i.e., $I$ and $S$ spin 
rf field strengths) obtained {on basis of a random initial pulse sequence} using the gradient-based optimal 
control algorithm GRAPE,\cite{oc2} the monotonic Krotov-based optimal control algorithm of Maximov et al.,\cite{dnp2} 
and the latter combined with smoothing as described in this paper. It is quite evident that the pulse sequences 
developed using the two former methods display quite significant oscillations, for which the appearance depends
 very much on the specific optimization. In the present case it looks like the GRAPE algorithm produces a more 
smooth, albeit still oscillating a lot in the "waves", pulse sequence than the Krotov-based algorithm. Slight 
variation in parameters, such as the number of steps in the waves, may radically change this pictures and in a 
later example we will see the opposite picture. The same variability will be seen upon inspection of the many 
optimal control NMR pulse sequences presented in literature so far. What is clear, however, is that the 
Krotov-based algorithm (as selected in this paper, but it could just as well be  GRAPE) combined with 
frequency-truncation smoothing leads to much smoother sequences which may easier be analysed analytically 
and which put less demands  on the spectrometer hardware upon practical implementation. 
{Apart from this the sequences display only relatively modest variations with respect to rf power consumption (root-mean-square (RMS) average rf
 powers (I and S spin values separated by /) of  141/71 Hz for GRAPE, 92/91 Hz for Krotov, and 12/102 Hz for smoothing Krotov) and are essentially identical with respect to coherence transfer efficiency (99\% of the theoretical maximum for all methods).}

{While rigid statements on the optimization speed and convergence require a much more thorough 
analysis (to be presented elsewhere), examples like the one in Fig. \ref{fig1} provide the following 
crude estimate. Smooth Krotov is computation-speed wise quite similar to GRAPE, while the original Krotov 
approach is somewhat faster (in the order of a factor 3-5, although this number highly depends  on the 
specific optimization). Out of 1000 optimizations based on random initial pulse sequences only 80-90 \% of 
the GRAPE and Krotov optimizations led to sequences with more than 90\% of the nominal transfer efficiency, 
while all sequences in this specific case passed this limit for the smooth Krotov approach. The great benefit 
of smooth Krotov is that it provides smooth pulse sequences, and the benefit of both Krotov variants is that 
they enable optimization with much more coarse time dicretization (longer and fewer pulses) than the GRAPE algorithm 
due to its fundamentally different optimization strategy.}

\begin{figure}
\includegraphics[scale=0.35,angle=-90]{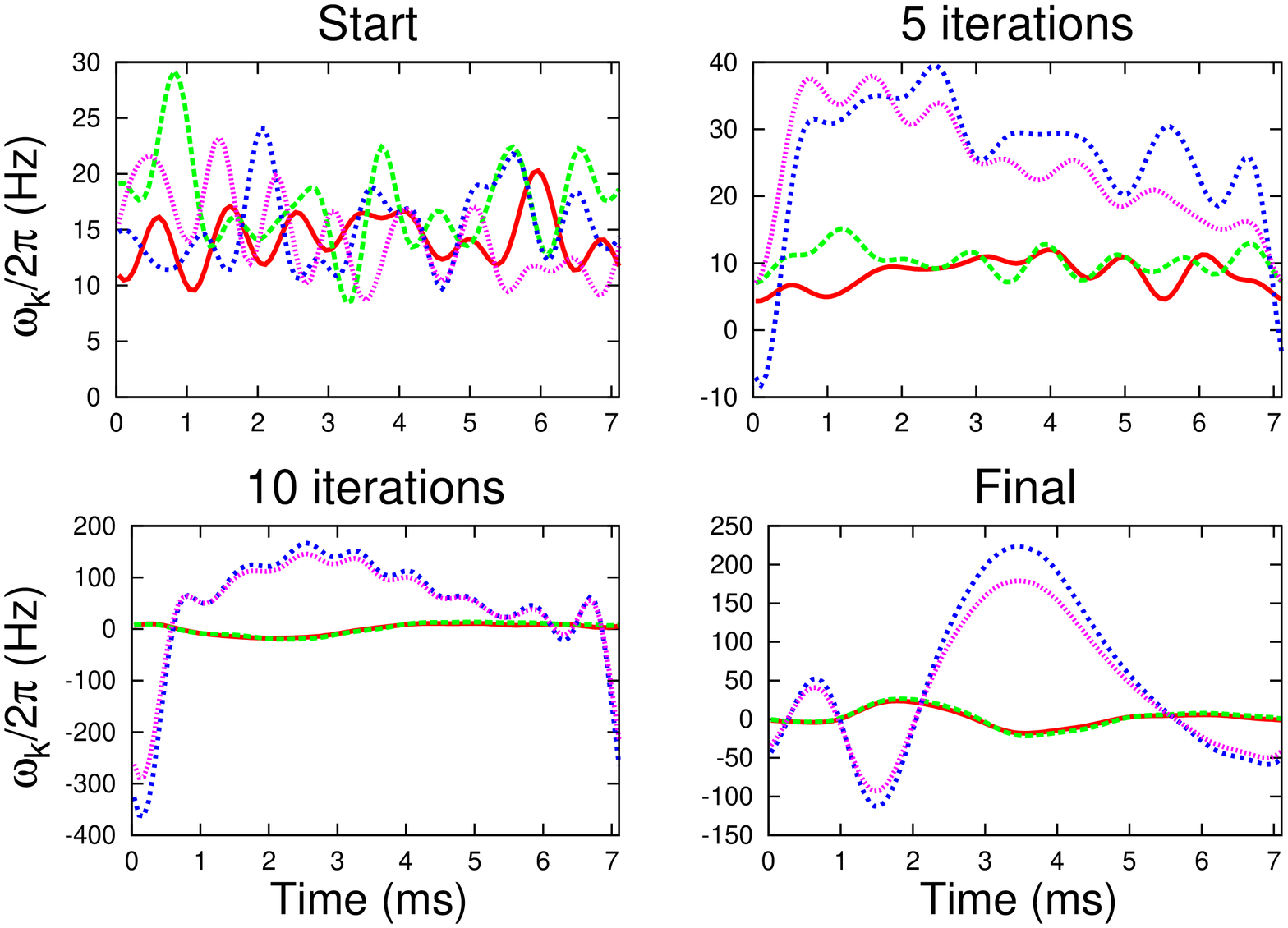}
 \caption{}
\label{fig2}
\end{figure}

\begin{figure}
\includegraphics[scale=0.35]{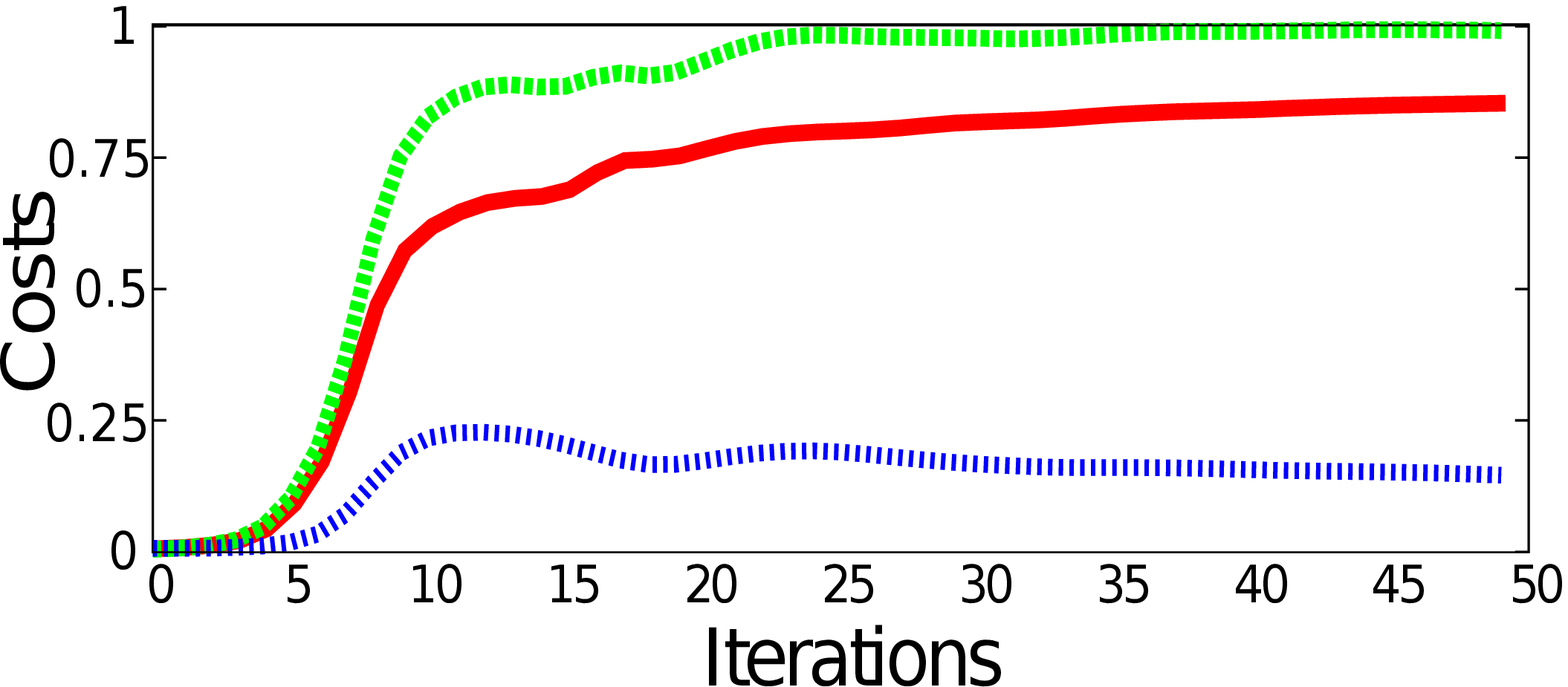}
 \caption{}
\label{fig3}
\end{figure}

Figure \ref{fig2} gives snapshots of pulse sequences throughout optimization using the smoothing Krotov-based 
optimal control algorithm starting out from a random pulse sequence (with maximum amplitude of 100 Hz). 
The snapshots illustrate gradual adoption of the optimal sequence to a smooth appearance. It is evident 
that already after 5 iterations, the control fields on the $I$ spin almost vanishes, leaving the major 
action to the $S$ spin control fields for the remaining optimization into the final pulse sequence. The 
final pulse sequence have quite low rf-power consumption, due to the reducing effect of the running cost  
($\lambda_1$ = $10^{-4}$ s$^{2}$rad$^{-2}$, $\lambda_2$ = $\lambda_1$ for both the $I$ and $S$ spins), and 
the rf fields vary smoothly as an effect of the smoothing algorithm. The progress of the cost function and 
its constituents (the overall functional (cost), the transfer efficiency, and the running cost (penalty)) 
throughout the optimization is illustrated in  Fig. \ref{fig3} which also provide a numerical demonstration 
of monotonic convergence. It is evident that the functional converges to a value which, when considering the 
subtractive term from the running cost, is equal to or close to the theoretical limit after a number of 
iterations where the efficiency (final cost) and penalty (running cost) have displayed some exchanging 
oscillations before converge to the optimal values. 



\subsection{Optimal control for satellite and central transition excitation in $^{23}$Na MRI}

Optimal control mediated experiment design is by no means restricted to spin-1/2 nuclei or systems of these.
 Many challenging optimization examples may be found for NMR spectroscopy or MRI in concern of quadrupolar nuclei,
 as recently demonstrated by optimal control pulse sequences for improved multiple-quantum magic-angle-spinning 
NMR\cite{ocmqmas}  and for selective excitation of the central transition of $^{23}$Na (spin $I$ = 3/2) in 
presence of residual quadrupole couplings for magnetic resonance imaging purposes.\cite{mri4,mri5,mri6}  The 
latter application may be very important for $^{23}$Na MRI of, e.g., cartilage where the sodium concentration
 may be considered a reporter for disorders and degradation.\cite{mri4,shapiro}  In such applications, the 
ability to separate  $^{23}$Na ions with large and very small (vanishing) residual quadrupolar couplings is 
regarded important as they represent different populations of relevant ions.

In this section, we demonstrate the use of the smoothing Krotov-based optimal control algorithm to design 
pulse sequences which selective excites the central transition ($-\frac{1}{2}$,$\frac{1}{2}$) or the 
satellite ($-\frac{3}{2}$,$-\frac{1}{2}$ or $\frac{1}{2}$,$\frac{3}{2}$) transitions of $^{23}$Na ions 
characterized by a residual quadrupole coupling frequency of $\omega_Q/2\pi$ = 60 Hz. 
In this case the size of the quadrupole coupling and the rf irradiation fields are comparable (often 
referred to as the intermediate regime) implying that experiment design by standard analytical means is 
not straightforward.

The optimization involves the secular, high-field-approximated first-order Hamiltonian 
\begin{equation}
H=\frac{\omega_Q}{2}(3I^2_z-I(I+1))+\omega_1(t)I_x+\omega_2(t)I_y,
\label{eq13}
\end{equation}
where $\omega_Q$ is quadrupole coupling frequency while  $\omega_1(t)$ and $\omega_2(t)$ are control rf
 fields corresponding to the spin operators $I_x$ and $I_y$, respectively (all frequencies in angular units). 
 The nature of the optimization will be a standard state-to-state transfer with the final cost expressed by
 Eq. (\ref{eq5}) with the destination operator $C$ representing $x$-phase coherence on either the central 
transition or the satellite transitions. The initial state corresponds to the thermal equilibrium $\rho_0=I_z$.

\begin{figure}
\includegraphics[scale=0.35, angle=-90]{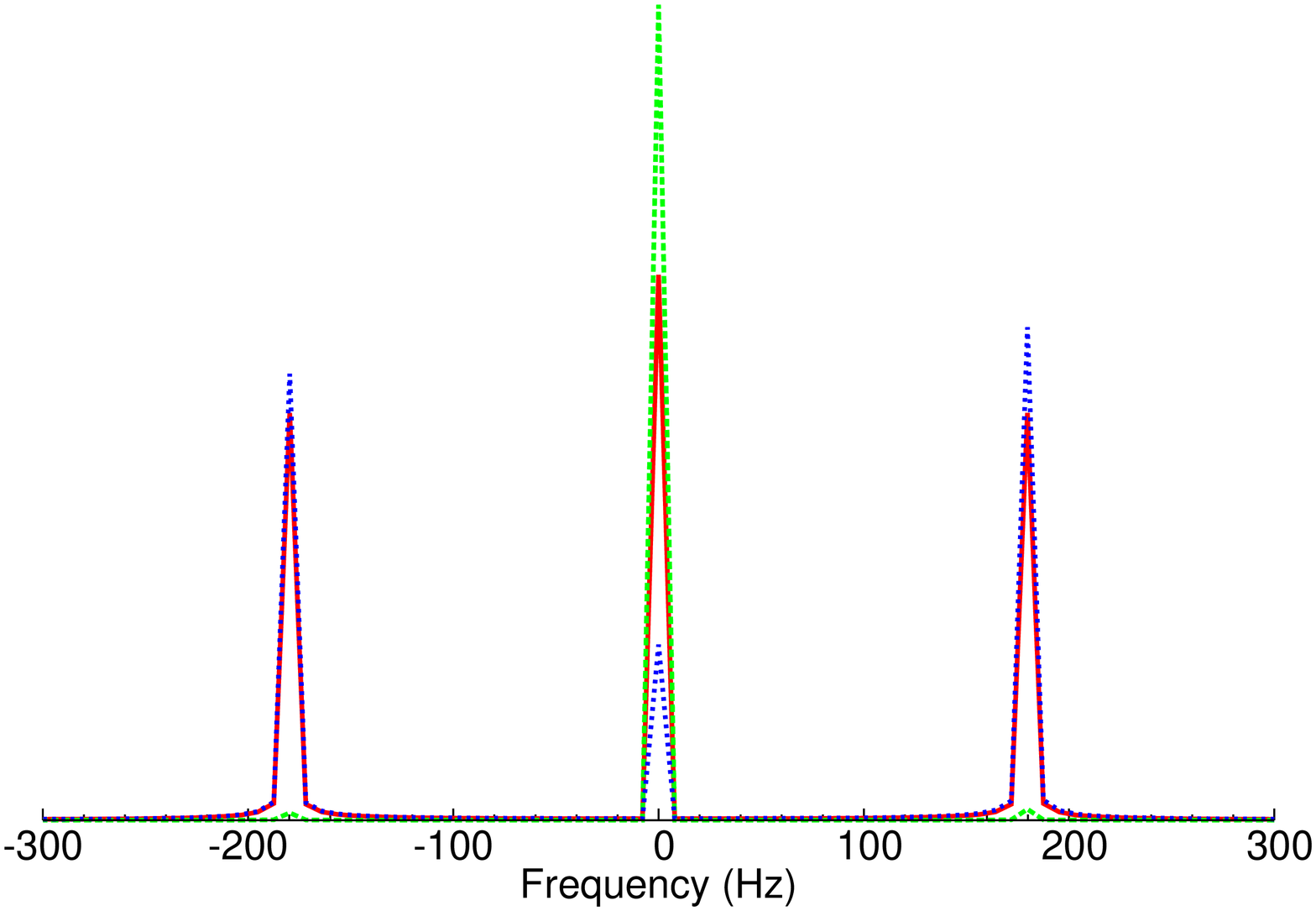}
 \caption{}
\label{fig4}
\end{figure}

Figure \ref{fig4} illustrates the performance of optimal control pulse sequences designed to enhance the central
 transition (in this case leading to efficient suppression of  the satellite transitions although not requested 
in the cost function used for the optimization) or excite the satellite transitions while suppressing the 
central transition (also here the suppression was not specified in the cost function) relative to a hard 
(infinitely strong) 90$^o$ pulse exciting both types of transitions.  The simulated spectra clearly reveal 
that  (i)  the central-transition selective excitation optimal control sequence offers a sensitivity enhancement, relative to standard single-pulse excitation, by the maximal factor of 1.5 as discussed previously by Jerschow and coworkers,\cite{mri2} while ensuring 
reasonable suppression of the satellite transitions and  (ii) the pulse sequence designed to suppress the 
central transition lead to satellites with unaltered intensity while the central transition is almost removed. 

\begin{figure}
\includegraphics[scale=0.35]{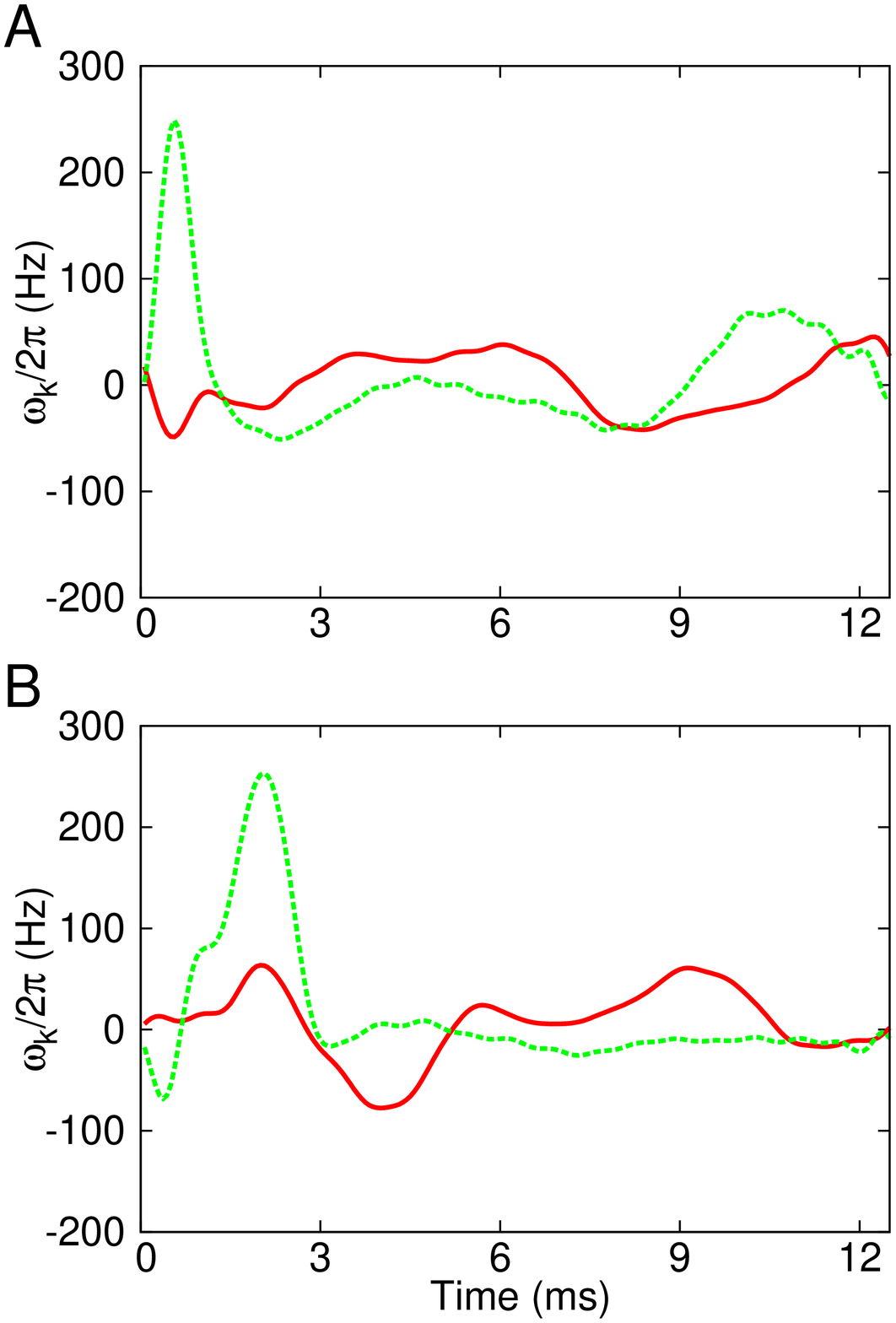}
 \caption{}
\label{fig5}
\end{figure}

The optimal control pulse sequences providing selective excitation of the central transition and the 
satellite transitions  are shown in Figs. \ref{fig5}A and \ref{fig5}B,  respectively. The sequences have a 
overall length of $2.25/f_Q$ (with the peak splitting $f_Q$  related to the quadrupole 
coupling as $ f_Q=3 \omega_Q/2\pi$) and accommodates 200 pulses of equal length.
{The central 
transition excitation pulse sequence is characterized by a RMS average rf power of 45 Hz and excites 
the central transition with an efficiency of 99\% and the satellites have been suppressed to
an efficiency of 1.4\% (percentages relative to the theoretical maximum). The satellite transition sequence is characterized by an RMS rf power of 54 Hz, with the satellites excited with an efficiency of 99\% and the central transition suppressed 72\%.} 

\begin{figure}
\includegraphics[scale=0.35]{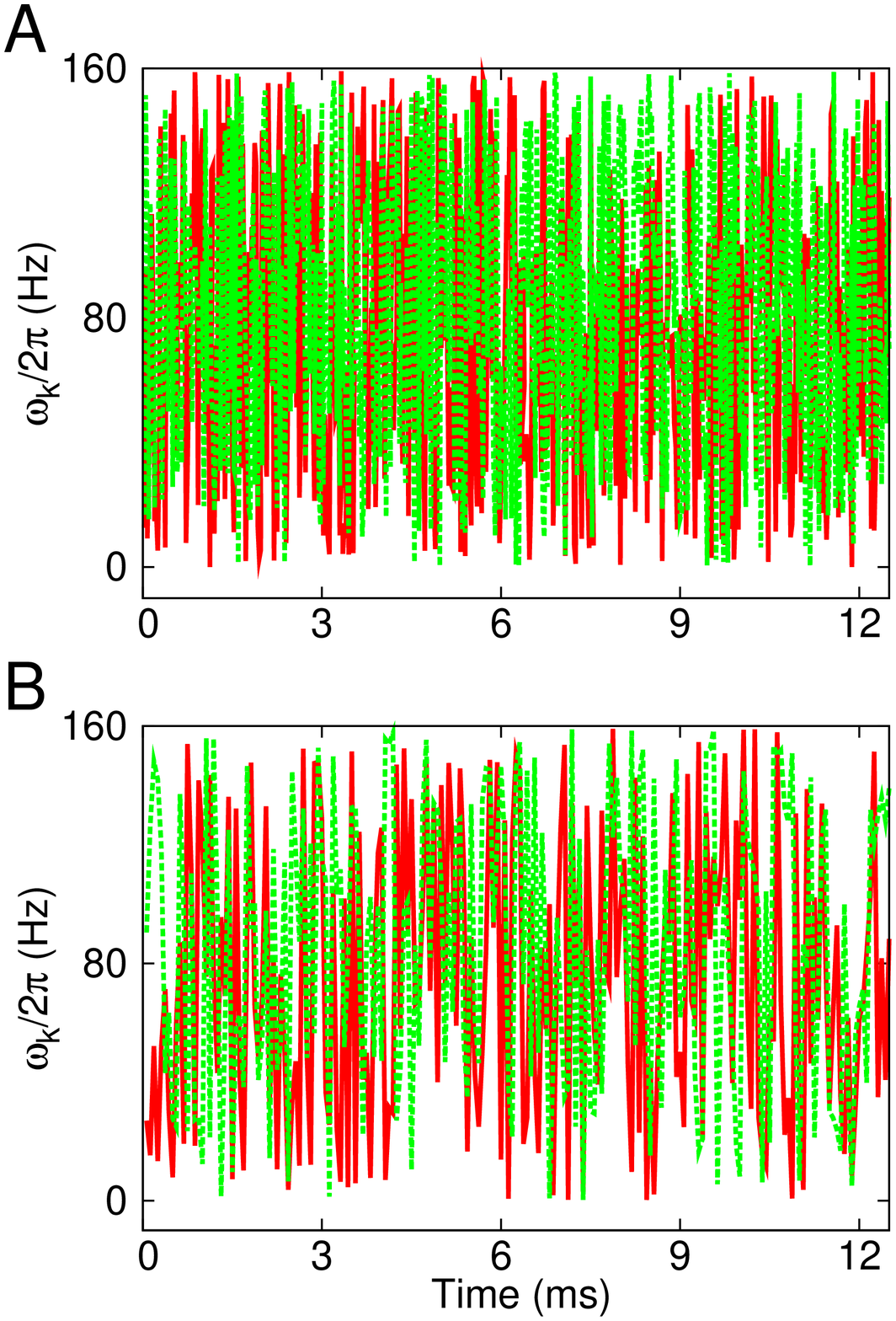}
 \caption{}
\label{fig6}
\end{figure}

It is clear from both sequences that the optimal pulse sequence is found in the regime  $\omega_{rf} \sim \omega_Q$ 
for which experiment design by standard analytical means is exceedingly difficult.  It is also evident that the 
smoothing algorithm used during optimization leads to smooth sequences offering easier implementation on conventional 
MRI instrumentation, than the corresponding sequences we obtained using previous optimal control formulations. The 
latter aspect is demonstrated in Fig. \ref{fig6}A by central transition-selective excitation sequences 
(corresponding to Fig. \ref{fig5}A) obtained using GRAPE  or in Fig.  \ref{fig6}B  using our previous Krotov-based
 optimal control software. The latter pulse sequences involved 500 pulses for GRAPE and 200 pulses for the 
Krotov-based approach to facilitate comparison with the pulse sequences earlier proposed by Jerschow and 
coworkers \cite{mri4}  for GRAPE. In the present implementation, the GRAPE-derived sequence displays much wilder 
oscillations than the Krotov-based sequence. 


\subsection{Broadband optimization}

In practical applications of optimal control theory for experiment design, it is often desirable to 
optimize the pulse sequences to be robust toward variation in one or more spin system parameters, 
introducing the concepts of broadband or bandselective excitation as described previously in the context 
of GRAPE  \cite{oc3} and Krotov-based \cite{dnp2}  methods. Such requests may readily be incorporated into 
our optimal control algorithm through a modified cost function
\begin{eqnarray}
\nonumber
J(\omega) &=& \kappa \mathrm{Tr}(U(T)U^{\dagger}(T))+\sum_{i}\mathrm{Tr}(C^{\dagger}U_i(T)\rho_0U^{\dagger}_i(T))\\
&-&\lambda\sum_k\int_0^T\omega_k^2(t)dt ,
\label{br1}
\end{eqnarray}
where each propagator $U_i(T)$ corresponds to a specific condition, such as different chemical shifts, 
couplings, or orientations of the sample relative to the magnetic field (in solid-state NMR), where $U$ is
 a common propagator of the system. 
The additional targets in the functional Eq. (\ref{br1}) corresponding to the broadband extension demands
 a modification of
the condition of a monotonic convergence and as concequence changing the form of Eq.(\ref{ap2}).
This leads to a modification of the target function $f_j(\omega'_j)$ in Eq.(\ref{ap2}) where the part 
involving propagators and Lagrange matrices should be replaced by a summation 
\begin{equation}
\dots\sum_{\mbox{conditions }i}A_iU'_{i,j-1}B_{i,j-1}A_i^{\dagger}\dots
\label{br2}
\end{equation}
with the propagator $U_i$ and adjoint matrix $B_i$ computed for each individual condition when the pulse 
sequence $\omega_k$ is the same for all propagators. 

\begin{figure}
\includegraphics[scale=0.35]{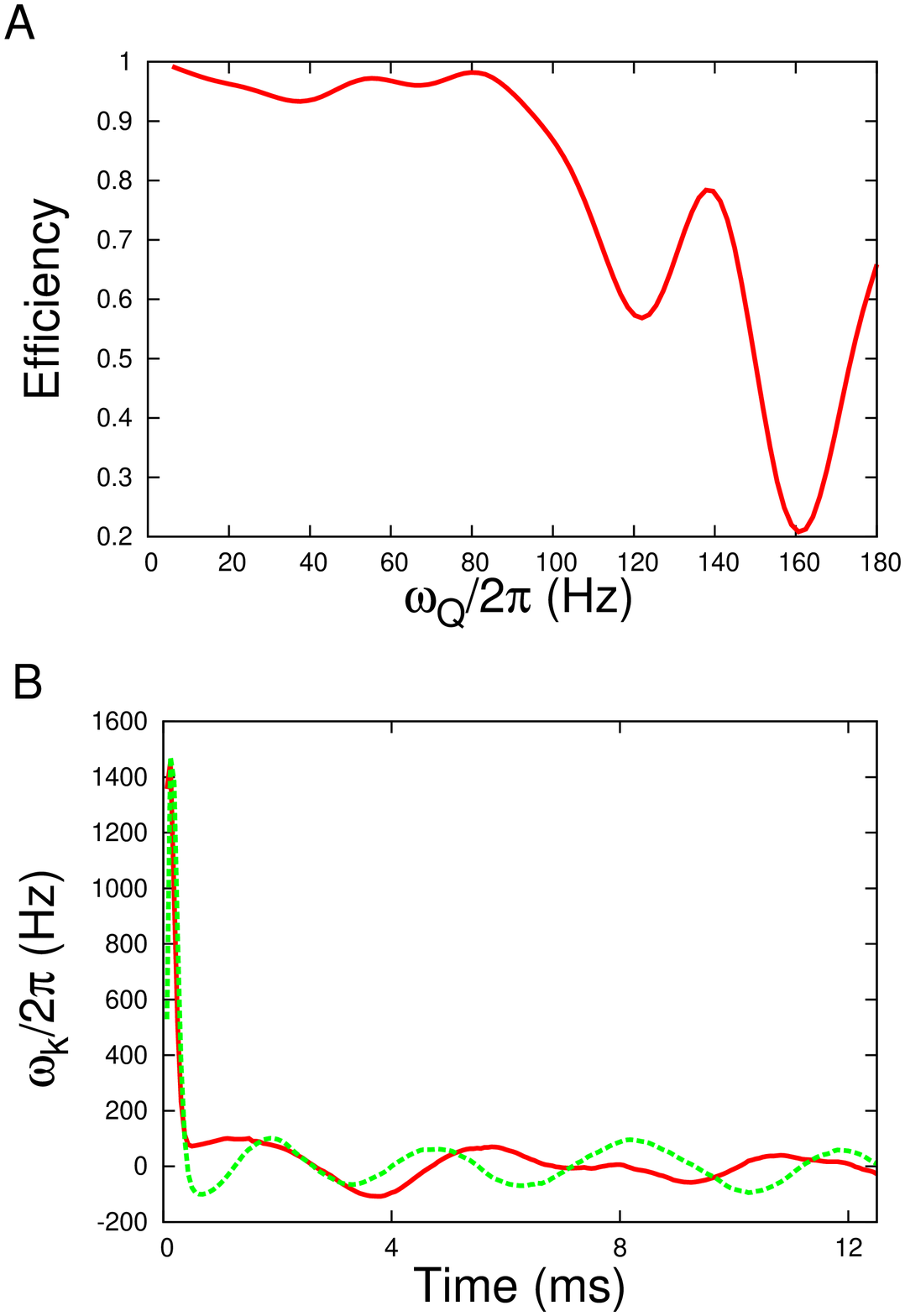}
 \caption{}
\label{fig7}
\end{figure}

To demonstrate the application of broadband optimization, we repeat  our optimization of $^{23}$Na MRI 
central transition enhancement experiments for a range  of quadrupole coupling around $\omega_Q/2\pi=60$ Hz
 with a deviation  of $\pm20$ Hz. This leads to a pulse sequence with a central transition excitation profile 
as function of the quadrupole coupling frequency as illustrated in Fig. \ref{fig7}, revealing very good 
excitation in proximity of $\omega_Q/2\pi=60$ Hz as requested in the optimization. We note that it is obviously
 possible to specify specific regions of excitation and no excitations by entering additional conditions in the 
target function and algorithm.
\\
 
\section{Conclusion}

In conclusion, we have presented a smoothing monotonically convergent optimal control algorithm for efficient 
design of pulse sequences for magnetic resonance applications. The proposed frequency-truncation algorithm has
 been incorporated into a Krotov-based optimal control procedure and demonstrated for a few of  NMR examples 
where it leads to much smoother optimal pulse sequences than obtained using previous methods. {We note that
 previous methods may eventually lead to smooth sequence by themselves, in particular for simple optimizations 
without inhomogeneities and broadband performance request and in cases where educated guesses are present to 
initiate the optimization. However, in the most general case optimal control typically leads to pulse sequences
 with significant oscillations in phase and amplitude. In such cases our new algorithm offers a remedy to
 incorporate smoothness in the optimization.} The smooth sequences improves the possibilities for obtaining 
theoretical insight for the optimal sequences and facilitates implementation on typical spectrometer hardware. 
 We anticipate that the methods will find widespread applications for design of experiments within liquid- and 
solid-state NMR spectroscopy, MRI, and dynamic nuclear polarization. 

\begin{acknowledgments}
This work was supported by the EU 6th Framework Program in terms of the BIODNP research infrastructure project.
 Support from the Danish National Research Foundation, the Danish Centre for Scientific Computing, and the Kavli 
Institute for Theoretical Physics, University of California, Santa Barbara (NCN) is acknowledged.  GT and JS were partially supported by the ``Agence Nationale de la Recherche'' (ANR),
Projet Blanc C-QUID number BLAN-3-139579.
We thank Christoffer Laustsen for discussion on $^{23}$Na MRI. 
\end{acknowledgments}

\newpage

\section{Figure Captions}

\vskip 1 cm

Fig. 1. (Color online) Pulse sequences for spin-state- and coherence-order-selective  $S^+ \rightarrow I^-S^{\alpha}$ 
coherence transfer in a $J$-coupled ($J$=140 Hz) heteronuclear two-spin system designed using optimal control based on 
the GRAPE algorithm (first row),  the original monotonic Krotov-based algorithm (second row), and the latter combined 
with frequency-truncation smoothing  (third row). The red and green lines correspond to $x$- and $y$-phase rf  control 
fields for the spins $I$ (left column) and $S$ (right column), respectively.

\vskip 1 cm

Fig. 2. (Color online) Representative pulse sequences obtained throughout the optimization in Fig. 1, 
illustrating the gradual adoption of the pulse sequence to provide optimum $S^+ \rightarrow I^-S^{\alpha}$ 
coherence transfer with modest rf-power consumption and smooth rf variation. The red and green lines correspond 
to $x$- and $y$-phase control fields for spin $I$ while blue and pink lines represent $x$- and $y$-phases of 
the $S$-spin control fields.

\vskip 1 cm

Fig. 3. (Color online) The dependence of the functional $J(\omega)$, the efficiency $\phi_2$, and the 
running cost (penalty) throughout the iteration steps of a smoothing monotonic Krotov-based optimal 
control design of the pulse sequence for $S^+ \rightarrow I^-S^{\alpha}$ coherence transfer, shown and 
analysed in Figs. 1 and 2. The red curve represents the functional $J(\omega)$ {(defined by Eqs. (4) 
and (16); for transparency, we have subtracted the correction terms in Eq. (16) which otherwise would add a 
constant contribution to the functional in all points),} the green the efficiency $\phi_2$, and blue the 
penalty cost (see text). {All ordinates have been scaled by a factor $Tr\{CC^\dagger\}$.}

\vskip 1 cm

Fig. 4. (Color online) Simulated $^{23}$Na NMR spectra of sodium with a residual quadrupole coupling 
frequency of $\omega_Q/2\pi$ = 60 Hz excited using a standard 90$^o$ hard pulse (red line), the optimal 
control pulse sequence in Fig. 5A designed to excite exclusively the central transition (green line), 
and the pulse sequence in Fig. 5B designed to excite the satellite transitions only (blue line).

\vskip 1 cm

Fig. 5. (Color online) Optimal control pulse sequences for $^{23}$Na MRI, which for the case of a residual 
quadrupole coupling of 60 Hz offer selective and enhanced excitation of the central transition {\bf(A)} 
or excitation of the satellite transition while suppressing the central peak of the $^{23}$Na NMR/MRI 
spectrum \textbf{(B)}. The red curve represents $x$-phase control fields while the green curve 
represents $y$-phase controls.  The length of the pulse sequences are 12.5 ms.

\vskip 1 cm

Fig. 6. (Color online) Optimal control pulse sequence for excitation of the central peak of
 the quadrupole spectrum of $^{23}$Na \textbf{(A)} (parameters as described in relation to Fig. 4)
 obtained using the GRAPE algorithm and \textbf{(B)} the original Krotov based algorithm. The red and green
 curves represent $x$- and $y$-phase control fields.  The length of pulse sequences in both cases is 12.5ms.

\vskip 1 cm

Fig. 7. (Color online) (a) Excitation profile (normalized) for the central transition as function of the 
quadrupole coupling frequency for a pulse sequence optimized using our smooth variant of the Krotov 
algorithm and specifying "broadband" excitation for a range of quadrupole coupling frequencies in 
proximity of 60 Hz. \textbf{(b)} The length of the pulse sequence is 12.5 ms,  {the RMS average rf power is 170 Hz,
 and the average transfer efficiency in the range of quadrupolar couplings between 40 and 80 Hz is 1.89 (corresponding to 95 \% of the theoretical maximum).}

\newpage

\vskip 1 cm
Figure 1:
\vskip 1 cm

\newpage

\vskip 1 cm
Figure 2:
\vskip 1 cm

\newpage
\vskip 1 cm
Figure 3:
\vskip 1 cm

\newpage
\vskip 1 cm
Figure 4:
\vskip 1 cm

\newpage
\vskip 1 cm
Figure 5:
\vskip 1 cm

\newpage
\vskip 1 cm
Figure 6:
\vskip 1 cm

\newpage
\vskip 1 cm
Figure 7:
\vskip 1 cm

\end{document}